\documentstyle[12pt]{article}
\author{P.B.Wiegmann\\
{\it James Frank Institute and Enrico Fermi Institute }\\ {\it of the
University of Chicago},
\\{\it 5640 S.Ellis Ave., Chicago IL, 60637}, \\ {\it and Landau Institute
for Theoretical Physics}\\ e-mail:wiegmann@control.uchicago.edu\\
\vspace{0.5cm}\\
A.V.Zabrodin\\
{\it Institute of Chemical Physics, Kosygina St. 4, SU-117334,}\\ {\it
Moscow, Russia,}\\
{\it and ITEP, Moscow, 117259, Russia},\\ e-mail: zabrodin@vxitep.itep.ru}
\title{ Algebraization of difference eigenvalue equations related to
$U_q(sl_2)$}
\date{}
\begin{document}

\maketitle
\begin{abstract}
A class of second order difference (discrete)
operators with a partial
algebraization of the spectrum is introduced. The eigenfuncions of the
algebraized part of the spectrum are polinomials (discrete polinomials).
Such difference operators can be constructed by means of $U_q(sl_2)$, the
quantum deformation of the $sl_2$ algebra. The roots of polinomials
determine the spectrum and obey the Bethe Ansatz equations. A particular
case of difference equations for $q$-hypergeometric and Askey-Wilson
polinomials is discussed. Applications to the problem of Bloch electrons in
magnetic field are outlined. \end{abstract}

\newpage

\section{Introduction}
In this paper we attempt to describe difference equations of second order
in one variable,
\begin{equation}\label{1}
a(z)\Psi(q^2z)+d(z)\Psi(q^{-2}z)+v(z)\Psi(z)=E\Psi(z), \end{equation}
having polynomial eigenfunctions of the form \begin{equation} \label{Psi}
\Psi(z)=\prod_{m=1}^{N}(z-z_m).
\end{equation}
Here $q$ is a parameter
and the polinomials are parametrized by their roots $z_m$. We call it
algebraization of the spectrum.

Difference equations are closely related to the discrete equations. The
discrete equation
\begin{equation}\label{21}
a_n\psi_{n+1}+d_n\psi_{n-1}+v_n\psi_n=E\psi_n\,. \end{equation}
may be obtained from the difference
equation (\ref{1}) by setting $z=q^{2n}$: \begin{eqnarray} \label{4}
\Psi(q^{2n})=\psi_n,\, a(q^{2n})=a_n,\,d(q^{2n})=d_n,\,v(q^{2n})=v_n\,.
\end{eqnarray}
Therefore, if the difference equation has a polynomial solution,
then one can find a solution of the corresponding discrete equation as a
polynomial on a discrete support: \begin{equation}
\label{3} \psi_n=\prod_{m=1}^{N}(q^{2n}-z_m). \end{equation}

Generally only a part of the spectrum of difference and discrete operators
can be
algebraized. However, in some cases the entire spectrum is algebraic. The
important examples of this type are periodic discrete equations. They
appear
if
$q$ is a root of unity, i.e., $q^{2Q}=1$, where $Q$ is an integer. Then we
obtain equations with periodic coefficients $ a_n=a_{n+Q},\, d_n=d_{n+Q},\,
v_n=v_{n+Q}$ and the periodicity condition
\begin{equation} \label{10}\psi_n=\psi_{n+Q}. \end{equation}

Some of the equations (\ref{1},\ref{21}) have important physical
applications. In particular, the method of algebraization has been applied
\cite{WZ,WZ1} to the problem of Bloch electrons in magnetic field on a
lattice (the Azbel-Hofstadter problem) \cite{Bemf}.

Discrete equations with periodic coefficients are reacher than the
difference equation with $|q|=1$. One can impose the quasiperiodic boundary
condition
\begin{equation}
\label{11}\psi_n=e^{ikn}\psi_{n+Q},
\end{equation}
where $k$ is the Bloch momentum. In this case the spectrum generally has
$Q$ bands. The periodic boundary condition (\ref{10}) describes the
crossection of the band spectrum at $k=0$. The quasiperiodic equations
($k\ne 0$) have no difference analog. We do not consider them in this
paper.

The class of algebraized operators may be extended by the "gauge"
transformation $\Psi(z)\rightarrow U(z)\Psi(z)$, where $U(z)$ may not be a
polynomial. Then $a(z)\rightarrow
U(q^2z)^{-1}a(z)U(z),\,d(z)\rightarrow U(q^{-2}z)^{-1}d(z)U(z),\,
v(z)\rightarrow v(z)$.
In particular, one can always choose the Jacobi operator (\ref{1},\ref{21})
to be symmetric: $a(z)=d(q^2 z)$.

In Appendix A we show that the difference equation (\ref{1}) has polynomial
solutions if
$a(z),\,d(z),\,v(z)$ are certain Laurent polinomials of order 2. They are
determined by 7 parameters (not including $q$) and may be computed
directly. However, the algebraization of difference equations has an
intimate relation with the representation theory. The relation between
representations of the $sl_2$ Lie algebra and second order differential
equations having polynomial solutions is known ( see
e.g.\cite{Wint,Gursey,ZU,BV} for early works, \cite{T1,Ush,T2} for recent
systematic treatment and \cite{Sh} for a review). In this paper we show
that difference equations having polynomial solutions can be classified
according to representations of the quantum algebra $U_q(sl_2)$, the
$q$-deformation of $sl_2$.

To construct linear operators with partially algebraized spectrum we employ
the following strategy. Consider the algebra ${\cal D}$ of difference
operators acting in the space of complex functions. Let ${\cal A}\subset
{\cal D}$ be its subalgebra such that ${\cal A}$ has a finite-dimensional
irreducible representation. This means that
elements of
${\cal A}$ leave invariant a finite-dimensional functional subspace. Some of
the
eigenfunctions of operators from ${\cal A}$ must then belong to this
subspace. Let us choose the invariant functional subspace to be the linear
space $Pol_n$ of polinomials of degree at most $n$. Our task, then, will be
to realize a finite dimensional representation of the algebra ${\cal A}$ in
this space.

The basis of the space $Pol_n$ may be choosen as monomials $z^k$,
$k=0,1,2,\ldots ,n$. In this basis, the algebra ${\cal A}$ is naturally
decomposed into the raising (lowering) parts ${\cal A}_{+}$ (${\cal
A}_{-}$) and diagonal operators ${\cal A}_{0}$ (Cartan subalgebra) with the
property ${\cal A}_{+}{\cal A}_{0}\subset {\cal A}_{+}$, ${\cal A}_{-}{\cal
A}_{0}\subset {\cal A}_{-}$. The space $Pol_n$ is invariant if and only if
${\cal A}_{+}z^{n}={\cal A}_{-}z^{0}=0$. This means that the representation
of ${\cal A}$ must have highest and lowest weights. Then
diagonalizable operators from ${\cal A}$ must have $n+1$ different
polynomial eigenfunctions. When the subalgebra ${\cal A}$ and its
representation are given we want to select the difference operators of
second order. In this case roots of the polynomial eigenfunctions satisfy
Bethe equations.

This
strategy has been used for construction and classification of differential
operators having polynomial eigenfunctions
\cite{Wint,Gursey,ZU,BV,T1,T2,Sh,Olver} (a different approach was suggested
in \cite{Ush}). Let
us use here the same notation ${\cal D}$ for the algebra of differential
operators. Then the subalgebra ${\cal A}\subset {\cal D}$ is a factor of
$U(sl_2)$ (the universal enveloping of $sl_2$) over its center. The algebra
$sl_2$ can be realized by first order differential operators,
\begin{equation} \label{Dmod}S_3=z{\frac{d }{dz}}-j,\,\,\,
S_{+}=z(2j-z{\frac{d }{dz}}),\,\,\, S_{-}={ \frac{d }{dz}}\,.
\end{equation} In the invariant subspace $Pol_{2j}$ the representation
(\ref{Dmod}) has highest and lowest weights $\pm j$ (integer or
half-integer spin of the representation). This realization provides the
embedding of ${\cal A}$ into the algebra of differential operators ${\cal
D}$. More precisely,${\cal A}$ in this approach is identified
with the factoralgebra $U(sl_2)/({\vec S}^{2}-j(j+1))$ over the ideal
generated by ${\vec S}^{2}-j(j+1)$, where ${\vec S}^{2}$ is the Casimir
(central) element.

Then the
Hamiltonian of the Euler top,
\begin{equation}
\label{Glim}H=\sum \alpha_{ij}S_iS_j+\sum \beta_iS_i\,, \end{equation}
i.e., a general bilinear form in the $sl_2$ generators gives a family
of second order differential operators having in general $2j+1$ independent
polynomial eigenfunctions.

Similar arguments may be applied to the quantum algebra $U_q(sl_2)$. In
this case one may identify ${\cal A}$ with a certain factor of $U_q(sl_2)$
over its center. Its highest and lowest weight representations (
$q$-analogs of (\ref{Dmod})) are known and presented in Sect.2. In this
paper we show that homogeneous bilinear and linear
forms in generators of $U_q(sl_2)$ ($q$-deformations of the Euler top
(\ref{Glim})) provide the difference equations of interest (Sect.3).

In addition to $U_q(sl_2)$
some other deformations of the $sl_2$ algebra
are known (Appendix B). They also can be used to construct difference
equations with polynomial solutions (we note that a
deformation not equivalent to $U_q(sl_2)$, which generates a subalgebra of
$U_q(sl_2)$,
has been used in the paper \cite{OT} in attempt to construct algebraized
difference operators (see Appendix B)). Among different deformations the
quantum algebra
$U_q(sl_2)$ plays a special role. It indicates a relation of difference
algebraized operators and classical discrete integrable systems (in fact
the former are
Lax operators for a class of nonlinear integrable equations). In this paper
we do not use the coproduct structure of the quantum algebra. However, we
anticipate an upgrading of the present approach
for quantum integrable systems.

Recently we have shown that the spectrum of the Harper equation (also
known as the discrete Mathieu or "almost Mathieu" equation) \begin{equation}
\label{5}\psi_{n+1}+\psi_{n-1}+2\cos(k+n\Phi)\psi_n=E\psi_n \end{equation}
at $\cos(kQ)=-1$, where $Q$ is the denominator of the parameter $\Phi=2\pi
P/Q$, is algebraized \cite{WZ1}: $$\psi_n=e^{-\frac{i}{2}\Phi
n(n-1)}\prod_{m=1}^{Q-1}(e^{i\Phi n}-z_m).$$ The roots $z_m$
obey the Bethe Ansatz like equations
\begin{equation} \label{6}z^2_l=
e^{i\pi P}\prod_{m=1,m\ne l}^{Q-1} {\ \frac{{e^{i\Phi}
z_l-z_m}}{{z_{l}-e^{i\Phi} z_m}}},\,\, \,l=1,...,N \end{equation}
and determine the spectrum \footnote{A more general algebraization in terms
of meromorphic functions on algebraic curves
rather than polinomials has been introduced by Faddeev and Kashaev
\cite{FK} for quasiperiodic Harper's equation for an arbitrary $k$.}:
$$E=(1-e^{i\Phi})\sum_{m=1}^{Q-1}z_m\,. $$ The Harper equation
describes the Bloch particle on a square lattice in magnetic field
(Azbel-Hofstadter problem). In this paper we write the Bethe Ansatz
equations for a general class of difference operators having polynomial
eigenfunctions. In addition, we present a solution for the Bloch
particle on the triangular lattice in Sect.4.

In Sect.2 we give main formulas related to $U_{q}(sl_2)$ and its
representations. In Sect.3 we consider the spectral problems for difference
operators appearing as linear and bilinear forms in generators of
$U_{q}(sl_2)$. They are difference (discrete) analogs of the algebraic
forms of known differential equations having polynomial solutions
\cite{T1,Ush,T2,Sh,Olver,Littlejohn Ann.di Math138 35 1984}. Besides,
we point out a class of difference spectral problems solvable in terms of
$q$-deformed classical orthogonal polinomials with discrete measure (big
$q$-Jacobi polinomials).
Possible applications to the problem of Bloch particle in magnetic field
are outlined in Sect. 4. In Sect. 5 the continuum ("classical") limit is
discussed.
A related class of
difference equations with solutions of the form of symmetric Laurent
polinomials is treated in Sect.6. The solutions of a particular case of
these equations
are Askey-Wilson polinomials.
\bigskip

\section{The quantum algebra $U_{q}(sl_2)$ and difference operators} \medskip

The algebra $U_q(sl_2)$ (a $q$-deformation of the universal enveloping of
the $sl_2$) is generated by the elements $A,B,C,D,$ with the commutation
relations \cite{KR,Skl2,D,J,RTF}
\begin{eqnarray}
&&AB=qBA,\,BD=qDB,
\nonumber\\
&&DC=qCD,\,CA=qAC,
\nonumber\\
&&AD=1,\,[B,C]={{A^2-D^2}\over{q-q^{-1}}}. \label{ABCD}
\end{eqnarray}
The deformation parameter $q$ may be considered as a formal variable. In
the classical limit $q\rightarrow 1$, the quantum algebra turns into the
universal enveloping of $sl_2$: ${(A-D)/(q-q^{-1})}\rightarrow S_3,\,
B\rightarrow S_{+},\,C\rightarrow S_{-}$.

The central element of this algebra is a $q$-analog of the Casimir operator
\begin{equation}
\label{casimir}\Omega =\frac{q^{-1}A^2+qD^2}{(q-q^{-1})^2}+BC. \end{equation}
As $q\rightarrow 1$,\,
$\Omega - 2(q-q^{-1})^{-2}$ tends
to ${\vec S}^2+1/4$.

The commutation relations (\ref{ABCD}) are simply another way to write the
intertwining relation for the $L$-operator: \begin{equation}
\label{RLL=LLR}R({u/v})(L(u)\otimes 1)(1\otimes L(v))=(1\otimes
L(v))(L(u)\otimes 1)R({u/v})
\end{equation}
with the trigonometric $R-$matrix
\begin{eqnarray}
\label{Rmat} R(u)&\!\!=&\!\!{\frac 12}(q+1)(u-q^{-1}u^{-1})+{\frac 12}
(q-1)(u+q^{-1}u^{-1})\sigma _3\otimes \sigma _3+\nonumber\\
&&+(q-q^{-1})(\sigma
_{+}\otimes \sigma _{-}+\sigma _{-}\otimes \sigma _{+}) \end{eqnarray}
satisfying the Yang-Baxter relation ($\sigma _j$ are Pauli matrices;
$\sigma _{\pm }=(\sigma _1\pm i\sigma _2)/2$). Generators $A,\,B,\,C,\,D$
are matrix elements of the $L$-operator
\begin{equation}
\label{L1} L(u)=\left[
\matrix{{{ukA-u^{-1}k^{-1}D}\over {q-q^{-1}}}&C\cr
B&{{ukD-u^{-1}k^{-1}A}\over {q-q^{-1}}}\cr}\right]. \end{equation}
Here $u$ is the spectral parameter and $k$ is an additional parameter
 ("rapidity" at the site). Note that the $R$-matrix is the $L$-operator in
the spin 1/2 representation. It is given by the same matrix
 (\ref{L1}) for $k=q^{1/2}$ with elements: $\,A=q^{{\frac 12}\sigma _3},
\,D=q^{-{\frac 12} \sigma _3},\,B=\sigma _{+},\,C=\sigma _{-}$.

Irreducible finite dimensional representations of dimension $2j+1$ can be
expressed in the weight basis, where $A$ and $D$ are diagonal matrices:
$A=diag\,(q^{j},...,q^{-j})$. An integer or halfinteger $j$ is spin of the
representation. The value of the Casimir operator (\ref{casimir}) in this
representation is
\begin{equation}
\label{casimir2} \Omega _{j}={\frac{{q^{2j+1}+q^{-2j-1}}}{{(q-q^{-1})^2}}}.
\end{equation}

These representations can be realized \cite{Skl2} by difference operators
acting in the linear space of polynomials $F(z)$ of degree $2j$. Let us
introduce
"shift" operators $T_{+}$ and $T_{-}$: $T_{\pm }F(z)= F(q^{\pm 1}z)$,
$T_{+}T_{-}=1$,
\begin{equation}
\label{inv1}
T_{\pm }z=q^{\pm 1}zT_{\pm },\,\,\,\,\,\,\,\,\, T_{\pm }z^{-1}=q^{\mp
1}z^{-1}T_{\pm }.
\end{equation}
Then
\begin{eqnarray}
\label{reprs}
&&A=q^{-j}T_{+},\,\,\,\,\,D=q^{j}T_{-},
\nonumber\\
&&B=z(q-q^{-1})^{-1}\left(q^{2j}T_{-}-q^{-2j}T_{+}\right) \nonumber\\
&&C=-z^{-1}(q-q^{-1})^{-1}\left(T_{-}-T_{+}\right). \end{eqnarray}
Then the lowest weight vector is $F_0 (z)=1$ whereas $F_{2j}(z)=z^{2j}$ is
the highest weight vector: $CF_0 =0\,,\,BF_{2j}=0$. The realization
(\ref{reprs}) is a smooth $q$-deformation of the representation of the
$sl_2$ algebra by first order differential operators (\ref{Dmod}).
The Casimir operator (\ref{casimir}) in this realization is equal
to the $c$-number $\Omega _j$ (\ref{casimir2}). This means that
(\ref{reprs}) actually
gives a representation of the factoralgebra $U_q(sl_2)/(\Omega -\Omega _j)$
over the two-sided ideal generated by $\Omega -\Omega _j$.

If $q$ is a root of unity there is, in addition, three parametric family of
finite dimensional representations having, in general, no lowest and no
highest weight \cite{Skl2}. Sometimes they are called cyclic (or
unrestricted) representations \cite{Roche}. They correspond to discrete
quasiperiodic equations (\ref{11}) which has no direct difference analog .
\bigskip

\section{Second order difference operators related to $U_q(sl_2)$ }\medskip

We call a {\it difference operator} in a variable $z$ any linear
combination of the form $\sum _{l} f_{l}(z)T_{+}^{l}$, where $f_{l}(z)$ are
rational functions of $z$ and the sum is finite. Such operators form the
algebra ${\cal D}$ mentioned in the Introduction. The general form of a
difference operator of second order then is (cf. (\ref{1})) $$
\Delta =f_{1}(z)T_{+}^{s}+f_{2}(z)+f_{3}(z)T_{-}^{s}\,, $$
where $s$ is an integer.

Second order difference operators appear as linear ($s=1$) or
bilinear($s=2$) forms in $U_q(sl_2)$ generators. In fact operators coming
from linear forms are equivalent to some bilinear forms due to isomorphism
of the algebras generated by $A,\, q^{1/2}BA,\, q^{1/2}CD,\, D$ and $A,B,C,
D$. Nevertheless, it is convenient to consider them separately. We begin
with linear forms.

\subsection{Linear forms}
Consider a linear form in the quantum algebra generators \begin{equation}
\label{L} L=aA+dD+(q-q^{-1})(bB+cC)\,,
\end{equation}
where $a$, $b$, $c$, $d$ are parameters. The diagonalization of this
operator in $2j+1$-dimensional representations of $U_q(sl_2)$ leads to the
difference equation
\begin{equation}
\label{Leq} (cz^{-1}+aq^{-j}-bq^{-2j}z)\Psi(qz)+(-cz^{-1}+dq^{j}+bq^{2j}z)
\Psi(q^{-1}z) =E\Psi(z).
\end{equation}
This equation has $2j+1$ polynomial solutions. Let us parametrize a
polynomial by its roots (\ref{Psi}). Pluging (\ref{Psi}) in (\ref{Leq}) and
dividing both sides by $\Psi(z)$ we get
\begin{equation}
\label{divided1} a(z)\prod_{m=1}^{N}{\frac{{q^{2}z-z_m}}{{z-z_m}}}+
d(z)\prod_{m=1}^{N}{\frac{{q^{-2}z-z_m}}{{z-z_m}}}=E \end{equation}
where we have denoted
\begin{equation}
\label{a(z)lin} a(z)=cz^{-1}+aq^{-j}-bq^{-2j}z\,, \end{equation}
\begin{equation}
\label{d(z)lin} d(z)=-cz^{-1}+dq^{j}+bq^{2j}z. \end{equation}

Assume that $c\ne 0$. Then two different cases are possible: (1) $b\ne 0,\,
c\ne 0$,
(2) $b=0,c\ne 0$ (the case $b=c=0$ is trivial).

Let us first consider the general case (1).
Assume that $\Psi(z)$ is nondegenerate, i.e., all $z_m$'s are different.
The l.h.s. of (\ref{divided1}) is a meromorphic function, whereas the
r.h.s. is a constant. To make them equal we must cancel all the
singularities of the l.h.s. They appear at singular points of $a(z)$, and
$d(z)$ (simple poles at $z=0$ and $z=\infty$) and at $z=z_m$. The singular
part at $z=0$ vanishes automatically (note that no one of $z_m$ 's can be
equal to 0). The residue at infinity is equal to $ bq^{2j-N}(1-q^{2N-4j})$.
Its vanishing determines the degree of the polynomial: $N=2j$. Comparing
the constant terms in the both sides of (\ref {divided1}) one finds the
energy spectrum: \begin{equation}
\label{linenergy1} E=aq^{j}+dq^{-j}-b(q-q^{-1})\sum_{m=1}^{2j}z_m.
\end{equation}
Annihilation of poles at $z=z_m$ gives the following Bethe-ansatz
equations: \begin{equation}
\label{linBE1} {\frac{{bz_{l}^{2}+dq^{-j}z_{l}-cq^{-2j}}}{{
bq^{-2j}z_{l}^{2}-aq^{-j}z_l-c}}}= -\prod_{m=1,m\ne
l}^{2j}{\frac{{qz_l-z_m} }{{z_l-qz_m} }}.
\end{equation}

In the case (2) the operator $L$ is triangular
in the basis of monomials. It includes only $A, D$ and $C$ (the generators
of the
Borel subalgebra). As a result the coefficients of $\Psi(z)$ can be
recursively determined. Therefore
there is only one polynomial solution
for each degree $N=1,...,2j$. One can see it from (\ref{divided1}). Its
l.h.s. is regular at $z=\infty$ hence there is no restriction on degree of
the polynomials:
\begin{equation}
\label{linBE2} {\frac{{dq^{j}z_l-c}}{{aq^{-j}z_l+c} }}= q^{N}\prod_{m=1,
m\ne l}^{N}{\frac{{qz_l-z_m}}{{z_l-qz_m} }}. \end{equation}
The energy spectrum is given by the simple formula \begin{equation}
\label{linenergy2} E_N =aq^{N-j}+dq^{j-N}. \end{equation}
The Bethe equations
(\ref{linBE2}) must have exactly one solution (modulo permutation of the
roots).
In section 3.2 we discuss a relation of the solutions with
$q$-orthogonal polynomials. \medskip

\subsection{Bilinear forms} \medskip

Consider a general bilinear form in $U_q(sl_2)$ generators: \begin{eqnarray}
\label{G}
G&\!\!=&\!\! aA^2+dD^2+(q-q^{-1})(c_{2}CA+b_{2}BD+b_{3}BA+c_{3}CD)+ \nonumber\\
&&+(q-q^{-1})^{2}(b_{1}B^{2}+c_{1}C^{2})\,, \end{eqnarray}
where $a,\,d,\,c_{i},\,b_{i}$ ($i=1,2,3$) are arbitrary parameters.

In the representation of spin $j$ it is
a difference operator:
\begin{equation}
\label{diffeq} G\Psi (z)=a(z)\Psi (q^{2}z)+d(z)\Psi (q^{-2}z)-v(z)\Psi (z),
\end{equation}
where
\begin{equation}
\label{a(z)}
a(z)=b_{1}q^{-4j+1}z^{2}-b_{3}q^{-3j}z+aq^{-2j}+c_{2}q^{-j}z^{-1}+
c_{1}q^{-1}z^{-2}\,,
\end{equation}
\begin{equation}
\label{d(z)} d(z)=b_{1}q^{4j-1}z^{2}+b_{2}q^{3j}z+dq^{2j}-c_{3}q^{j}z^{-1}+
c_{1}qz^{-2}\,,
\end{equation}
\begin{equation}
\label{v(z)}
v(z)=(q+q^{-1})(b_{1}z^{2}+c_{1}z^{-2})+(c_{2}q^{-j}-c_{3}q^{j})z^{-1}
+(b_{2}q^{-j}-b_{3}q^{j})z\,.
\end{equation}
The difference operator given by the linear form (\ref{Leq}) is a
particular case of the bilinear form $$
aA^{2}+dD^{2}+(q-q^{-1})(q^{j}cCA+q^{j}bBD+q^{-j}bBA+q^{-j}cCD)\,, $$
where $q$ is to be
changed to $q^2$.

Let us plug (\ref{Psi}) in (\ref{diffeq}) and divide both sides by $\Psi(z)
$. We get
\begin{equation}
\label{res} a(z)\prod_{m=1}^{N} {\frac{{q^{2}z -z_m}}{{z- z_m}}}+
d(z)\prod_{m=1}^{N} {\frac{{q^{-2}z -z_m}}{{z-z_m}}} -v(z)=E\,.
\end{equation}

Suppose that at least one of the coefficients $c_{1},\,c_2,\,c_3$ is
nonzero, then there are three main different cases:

(i) at least one of $b_1,\,b_2,\,b_3$ is nonzero and both $a(z)$ and $d(z)$
(see (\ref{a(z)}), (\ref{d(z)})) are nonzero (this is the case of generic
position; the other two can be considered as exceptional cases);

(ii) All $b$'s are zero: the quadratic form (\ref{G}) includes only $
A,\,D$ and $C$ (generators of the Borel subalgebra of $U_q(sl_2)$). In this
case the eigenfunctions are
big $q$-Jacobi polynomials \cite{AW}. They include all $q$-deformed
classical orthogonal polynomials with discrete measure \cite{AW};

(iii) One of the functions $a(z)$ or $d(z)$ is identically zero. In this
case the second-order difference equation (\ref{diffeq}) reduces to a
first-order equation.

Let us consider the case (i) first. The l.h.s. of (\ref{res}) is a
meromorphic function, whereas the r.h.s. is a constant. To make them equal
we must cancel all the singularities of the l.h.s. They appear at singular
points of $a(z)$, $d(z)$ and $v(z)$ (double and simple poles at $z=0$ and $
z=\infty $) and at $z=z_m$ (again, we consider non-degenerate case when all
of them are simple poles). Note that the case when at least one of $z_m$'s
is zero (i.e., coincides with the pole of $a(z)$ and $d(z)$) needs special
consideration. At the moment we assume that $z_m\ne 0$. The singular part
at $z=0$ vanishes automatically. Vanishing of the singular part at
$z=\infty $\,, \begin{eqnarray}
\label{singinf}
&&b_{1}(q^{2N-4j+1}+q^{-2N+4j-1}-q-q^{-1})z^{2}+ b_{2}(q^{-2N+3j}-q^{-j})z+
\nonumber\\
&&b_{3}(q^{j}-q^{2N-3j})z+
z(q-q^{-1})b_{1}(q^{2N-4j}-q^{-2N+4j})\sum_{m=1}^{N}z_m\,, \end{eqnarray}
determines degree of the polynomial: $N=2j$. If $b_1 =0$ and
$b_{2}/b_{3}=-q^{2M}$ for an integer $M>-j$, $N=j+M$ is also possible.
Below we consider only the generic case $N=2j$.

Comparing the constant terms in the both sides of (\ref{res}) we find the
energy spectrum:
\begin{eqnarray}
\label{energy}
E&=&b_{1}(q-q^{-1})(q^{2}-q^{-2})\sum_{n<m}^{2j}z_{n}z_{m}- \nonumber\\
&&-(q-q^{-1})(b_{2}q^{-j+1}+b_{3}q^{j-1})\sum_{m=1}^{2j}z_m+
aq^{2j}+dq^{-2j}\,.
\end{eqnarray}
Finally, annihilation of poles at $z=z_m$ gives the following Bethe-Ansatz
equations
\begin{equation}
\label{BE1}{\frac{{d(z_l)}}{{a(z_l)}}}=q^{4j}\prod_{m=1,m\ne l}^{2j}{\ \frac{
 {q^2z_l-z_m}}{{z_l-q^2z_m}}},\,\,\,l=1,...,2j. \end{equation}
The Bethe equations is a system of $2j$ algebraic equations. It must have
exactly $2j+1$ solutions corresponding to different eigenfunctions. In the
case (i) all of them are polynomials of one and the same degree $2j$.

Note that $d(z)/a(z)$ is a rational function having at most 4 zeros and 4
poles. Therefore (\ref{BE1}) looks like a system of Bethe equations for a
$XXZ$-spin chain on at most 4 sites with different spins at the sites. The
reader familiar with the algebraic Bethe ansatz \cite{F,Gaudin}
should notice that eq.(\ref{res}) is the Baxter identity for eigenvalues
of the transfer matrix $t(z)$ for this system
\begin{equation}
\label{transfer}
\Psi (z)t(z)=a(z)\Psi (q^{2}z)+d(z)\Psi (q^{-2}z), \end{equation}
In this context the eq.(\ref{transfer}) determines $t(z)$ and $\Psi(z)$
provided that $a(z)$ and $d(z)$ are known functions.

In the special case $q^{2j+1}=1$ the difference operator has finite
dimension $2j+1$. Therefore polynomial egenfunctions and the Bethe Ansatz
cover all the spectrum. There are no more solutions other than polynomials.
\medskip

\subsection{Triangular and first order operators}\medskip

\subsubsection{$q$-Hypergeometric equations}

The difference operators that include only elements
$A,\,D$ and $C$ (case (ii))
lead to $q$-hypergeometric equations. These operators preserve not only the
space $Pol_{2j}$ for some $j$, but all the spaces $Pol_n$ for any $n \ge
0$. Indeed, they are lower triangular in the basis of monomials $z^k$.
The l.h.s. of (\ref{res}) is now regular at $z=\infty$ , so there is no
restriction on degree of the polynomials. The Bethe equations are valid for
any $N<2j+1$:
\begin{equation}
\label{BE2}{\frac{{d(z_{l})}}{{a(z_{l})}}}= q^{2N}\prod_{m=1,m\ne l}^{N} {\
\frac{{q^{2} z_l-z_m}}{{z_{l}-q^{2} z_m}}},\,\, \,l=1,...,N. \end{equation}
Triangularity leads to a very simple structure of the spectrum of $G$:
\begin{equation}
\label{Esolv} E_{N}=aq^{2N-2j}+dq^{2j-2N},\,\,\,\, N=0,...,2j. \end{equation}
A general operator of this kind has the same spectrum as the diagonal
operator $aA^{2}+dD^{2}$; it does not depend on
the coefficients $c_1, c_2, c_3$.

For the sake of completeness let us identify the solutions of
(\ref{diffeq}) in the case (ii) with known $q$-orthogonal polynomials.
Consider the big $q$-Jacobi
polynomials $P_n^{(\alpha ,\beta )}(z;\gamma ,\delta ;q^2)$ (where we have
used the standard notation from the textbook \cite{AW}; to shorter the
notation we
denote them as $P_n(z)$). They can be expressed in terms
$q$-hypergeometric
series:
\begin{equation} \label{q-hyp1}
P_n^{(\alpha ,\beta )}(z;\gamma ,\delta ;q^2)=_3\phi _2\left[ \begin{array}{cr}
q^{-2n},q^{2\alpha +2\beta +2n+2},~q^{2\alpha +2}\gamma ^{-1}z & \\ &
;q^2,q^2 \\
q^{2\alpha +2},-q^{2\alpha +2}\delta \gamma ^{-1} & \\ &
\end{array}
\right]
\end{equation}
and obey the difference equation \cite{AW}
\begin{equation}
\label{bigJac}
\begin{array}{c}
(q^{2\alpha -2\beta +2}+(q^{2\alpha }\delta -q^{2\beta }\gamma
)z^{-1}-\gamma \delta q^{-2}z^{-2})P_n(q^2z)+ \\ +(1+(\delta -\gamma
)z^{-1}-\gamma \delta z^{-2})P_n(q^{-2}z)- \\ -[((1+q^{2\alpha })\delta
-(1+q^{2\beta })\gamma )z^{-1}-\gamma \delta (1+q^{-2})z^{-2}]P_n(z)= \\
=(q^{-2n}+q^{2n+2\alpha +2\beta +2})P_n(z)\,. \end{array}
\end{equation}
This equation is equivalent to (\ref{diffeq}) provided $d\ne 0$,
$b_1=b_2=b_3=0$ and
$
\gamma -\delta =c_3d^{-1}q^{-j}\,,-\gamma \delta =c_1d^{-1}q^{1-2j}\,,
q^{2\alpha }\delta -q^{2\beta }\gamma =c_2d^{-1}q^{-3j}\,,q^{2\alpha
+2\beta }=ad^{-1}q^{-2-4j}\,. $

Note that (\ref{q-hyp1}) is formally symmetric with respect to the change
$n \leftrightarrow -\alpha -\beta -n-1$. This symmetry becomes important if
$m \equiv \alpha + \beta +n+1$ is a negative integer. Then the series
(\ref{q-hyp1}) truncates at $n$-th term if $n<|m|$, but if $|m|<n$ it
truncates at $|m|$-th term. One has $\alpha +\beta =\log _{q^2}(a/d)-2j-1$.
In particular, if $a=d$ the series (\ref{q-hyp1}) is symmetric with respect
to $n\leftrightarrow 2j-n$. In other words, for integer $j$ it gives only
$j+1$ different eigenfunctions (they are polynomials of degrees
$0,\,1,\,...,j$). It can be easily seen that in this case the original
operator contains Jordan cells and, therefore, is not completely
diagonalizable (i.e., it has less than $2j+1$ different eigenfunctions). In
this case the formula (\ref{q-hyp1}) still gives all the eigenfunctions.

At $\delta =0$, the big $q$-Jacobi polynomials $P_n^{(\alpha ,\beta
)}(z;\gamma ,0;q)$ reduce (up to an $n$-independent factor) to the little
$q$-Jacobi polynomials
$p_{n}^{(\beta ,\alpha )}(z\gamma ^{-1},q)$. They provide eigenfunctions of
the linear form
(\ref{L}) at $b=0$. In this case $
q^\beta =-1\,,q^\alpha =-ad^{-1}q^{-1-2j}\,,\gamma =cd^{-1}q^{-j} $.
It is known \cite{AW} that the
little $q$-Jacobi polynomials
can be expressed through
the $q$-hypergeometric function
${}_2\phi _1$:
\begin{equation} \label{q-hyp2}
p_n^{(\beta ,\alpha )}(z\gamma ^{-1},q)=_2\phi _1\left[ \begin{array}{cr}
q^{-n},q^{\alpha +\beta +n+1} & \\
& ;q,q\gamma ^{-1}z \\
q^{\beta +1} & \\
&
\end{array}
\right].
\end{equation}
Again, if the operator is not completely diagonalizable, (\ref{q-hyp2})
gives less than $2j+1$ different eigenfunctions.

The following comment is in order. In contrast to the linear forms
(\ref{linBE2}) the system of algebraic equations (\ref{BE2}) generally has
more than one solution for a given $N$. However, only one of them
corresponds to an
eigenfunction. The point is that the Bethe equations have been derived
under assumption that
all roots are nonzero. In fact
one or more zero roots are
not forbidden (for instance, $q$-Legendre polynomials of odd degree are odd
functions).
However, they change the residues at $z=0$ of the first two terms in
(\ref{res}), so this case needs a
special consideration. To
illustrate this let us give an example.

{\it Example}. Consider the spectral problem for the following quadratic forms:
$$
G_{1}=A^{2}-D^{2}+(q-q^{-1})(q^{j-1}CA+BD-BA-q^{1-j}CD), $$
$$
G_{2}=q^{-j}A^{2}+q^{j}D^{2}+(q-q^{-1})(q^{-j}CA+q^{j}CD)+ (q-q^{-1})^{2}C^{2}.
$$
Though $G_{1}$ corresponds to the case (i) and $G_2$ to the case (ii) they
lead to {\it identical} Bethe equations. For instance, at $j=1/2$ the
equation is
$$
{\frac{{qz_{1}^{2}-q^{1/2}z_{1}+q^{1/2}}}{{
q^{-1}z_{1}^{2}+q^{-1/2}z_{1}+q^{-1/2}}}}=q. $$
It has two roots:
$
z_{1}^{\prime}=0$ and $,z_{1}^{\prime\prime}=2(q^{1/2}-q^{-1/2})^{-1}. $
Both of them give an eigenfunction of $G_1$, while for $G_2$
$z_{1}^{\prime}$ is an artifact.)

\subsubsection{First order operators}

At last, we consider the case (iii).\footnote{ These operators appear in
harmonic
analysis on the quantum group \cite{K},
\cite{NM}.}
When, say, $d(z)=0$ the operator is
\begin{equation}
\label{twistprim} G_{0}=aA^{2}+(q-q^{-1})(c_{2}C+b_{3}B)A\,. \end{equation}

The Bethe equations (\ref{BE1}) are simplified as $$
a(z_l)\prod_{k=1, k\ne l}^{2j}(q^{2}z_{l}-z_{k})=0\,, $$ so eigenfunctions
and the spectrum can be easily found.
Let $z_{\pm}$ be the two roots of the quadratic equation $a(z)=0$,
then the $m$-th eigenfunction ($m=0, 1, ... , 2j$) and the spectrum are $$
\Psi_{m}(z)=\prod_{k=0}^{2j-1-m}(z-z_{+}q^{-2k})\prod_{l=0}^{m-1}
(z-z_{-}q^{-2l})\,,
$$
$$
E_{m}=b_{3}q^{-j}(z_{-}q^{2j-2m}+z_{+}q^{2m-2j})\,. $$

\subsubsection{ Algebraic structure unifying triangular and first order
operators}

The similarity of the spectra of operators of types (ii) and (iii) is not
an accident.
It turns out that these operators form a simple quadratic algebra.
Consider two operators of the type (iii) \begin{equation}
\label{H1}H_1=\mu _1 BA+\mu _2CA+\mu _3A^2\,, \end{equation}
\begin{equation}
\label{H2}H_2=\nu _1 DB+\nu _2 DC+\nu _3 D^2 \end{equation}
with arbitrary coefficients $\mu _i$ and $\nu _i$ and a general operator of the
type (ii)
\begin{equation}
\label{H3}H_3=(q^2-q^{-2})
\left({\frac{{\mu _1\nu _2 A^2+\mu _2\nu _1 D^2}}{{(q-q^{-1})^2}}}-q\mu
_2\nu _2C^2-\mu _2\nu _3DC-\mu _3\nu _2CA\right).
\end{equation}
It is straightforward to check that these operators form the closed algebra
\footnote{A similar algebra has been considered by Granovskii and Zhedanov
in \cite{GZh}, a particular case $h_1=h_2=h_3=0$ has been discussed in
\cite{Odes}, \cite{Fairlie}.}:
\begin{equation}
\label{algH}
q^{-1}H_{i}H_{j}-qH_{j}H_{i}=g_{k}H_{k}+h_k \end{equation}
where $\{ijk\}$ stands for any cyclic permutation of $\{123\}$. The
structure constants $g_i$ and $h_i$ may be expressed in terms of
coefficients $\mu _i$ and
$\nu_i$ and the value of the Casimir operator $\Omega$ (\ref{casimir}).

The algebra (\ref{algH}) allows one to find the spectrum of $H_1$, $H_2$ or
$H_3$ in an algebraic way \cite{GZh1}.
Suppose one knows an eigenvector of $H_1$. Then, some
linear combination of $H_i$ is a creation operator: it creates a new
eigenvector by acting to
the known one. As a result the spectrum of $H_1$ is found to be a trigonometric
function of the number of level. \bigskip

\section{Difference periodic equations and the group of magnetic
translations} \bigskip

Difference periodic equations and discrete equations with periodic coefficients
appear
when $q$ is a root of
unity:
\begin{equation}
\label{q}q=\exp(i\pi P/Q),
\end{equation}
where $P$ and $Q$ are coprime integers. Some of them have important
applications in physics.

Let us briefly recall the problem of Bloch particles in magnetic field
(Azbel-Hofstadter problem). Consider a particle on a two-dimensional square
lattice.
The Schr\"odinger equation has the form
\begin{equation}
\label{1000}\sum_{\vec\mu}t_{\vec\mu}e^{iA_{\vec\mu}(\vec n)}\psi (\vec
n+\vec\mu)=E\psi(\vec n)\,,
\end{equation}
where $\vec n$ is a lattice site, $\vec\mu$ is a lattice vector
and $ t_{\vec\mu}=t_{-\vec\mu}$ is a hopping amplitude. The gauge potential
$A_{\vec\mu}(\vec n)=-A_{-\vec\mu}(\vec n-\vec\mu )$ describes a
homogenious
magnetic field with flux $\Phi = 2\pi P/Q$ per plaquette of the square
lattice: $$
\prod_{plaquette}e^{iA_{\vec\mu}(\vec n)}=e^{i\Phi}\equiv q^2. $$
In this context the parameter $q$ has a clear interpretation: it is a flux
per half of the plaquette. The problem may be expressed in terms of
magnetic translations \cite{Bemf}:
$$
T_{{\bf \mu}}\psi(\vec n)=e^{iA_{{\bf \mu}}(\vec n)}\psi(\vec n+\vec\mu)\,. $$
Let ${\bf x}$ and ${\bf y}$ be unit vectors along $x$ and $y$ directions on
the square lattice and
$\vec n= (n_{x},n_{y})$.
They generate the algebra of magnetic translations \begin{eqnarray}
\label{tr1}
&&T_{\vec n}T_{\vec m}=q^{n_ym_x-n_xm_y}T_{\vec n +\vec m},\,\,\,\,\,
\,T_{\vec n}^{-1}=T_{-\vec n}\,,\nonumber\\ &&T_{{\bf y}}T_{{\bf
x}}=q^{2}T_{{\bf x}}T_{{\bf y}},\,\,\, T_{{\bf y}}T_{-{\bf
x}}=q^{-2}T_{-{\bf x}}T_{{\bf y}}\,. \end{eqnarray}
The operators $T_{\vec\mu}^Q$ are central elements. As an algebra over its
center the algebra of
magnetic
translations
is the finite Heisenberg-Weyl algebra.

In these terms the Hamiltonian of the problem is \begin{equation}
\label{1001} H=\sum_{\vec\mu}t_{\vec\mu}T_{\vec\mu}. \end{equation}

The quantum algebra $U_q(sl_2)$ (\ref{ABCD}) may be expressed through the
Heisenberg-Weyl generators, i.e., in terms of magnetic translations
\cite{WZ}. There are many different ways to do it (see e.g. \cite{BS} ).
One of them (for odd $P$) is
\begin{equation}
\label{realiz}
\begin{array}{l}
A^2=T_{
{\bf x}+{\bf y}},\,\,\,\,\, D^2=T_{-{\bf x}-{\bf y}},\,\,\,\,\,
CA=iq^{1/2}(q-q^{-1})^{-1}(T_{-
{\bf x}}+T_{{\bf y}}), \\ \\
BD=iq^{1/2}(q-q^{-1})^{-1}(T_{
{\bf x}}+T_{-{\bf y}}) ,\,\,\,\,\,
CD=iq^{-1/2}(q-q^{-1})^{-1}(T_{-
{\bf x}}+T_{-2{\bf x}-{\bf y}}), \\ \\
BA=iq^{-1/2}(q-q^{-1})^{-1}(T_{
{\bf x}}+T_{2{\bf x}+{\bf y}}),\,\,\\ \\
B^2=-(q-q^{-1})^{-2}(T_{{\bf x}-{\bf y}}+T_{3{\bf x}+{\bf
y}}+(q+q^{-1})T_{2 {\bf x}}),\\ \\
C^2=-(q-q^{-1})^{-2}(T_{-
{\bf x}+{\bf y}}+T_{-3{\bf x}-{\bf y}}+(q+q^{-1})T_{-2{\bf x}}). \end{array}
\end{equation}
The central element (\ref{casimir}) in this representation is
\begin{equation}
\label{Omega1} \Omega =-2(q-q^{-1})^{-2}. \end{equation}
Comparing with (\ref{casimir2}) we see that this value corresponds to the
representation (\ref{reprs}) with $j=(Q-1)/2$ and dimension $Q$. The same
value (\ref{Omega1}) corresponds to a 2-parametric family of the $Q$
-dimensional cyclic representations \cite{Skl2}.

Pluging (\ref{realiz}) into (\ref{G}) we obtain \begin{equation}
\label{Gtr}
\begin{array}{c}
G=aT_{{\bf x}+{\bf y}}
+dT_{-{\bf x}-{\bf y}}
-b_1T_{{\bf x}-{\bf y}}
-c_1T_{-{\bf x}+{\bf y}}-b_1T_{3{\bf x}+{\bf y} }-c_1T_{-3{\bf x}-{\bf y}}+\\
+iq^{1/2}(c_2T_{{\bf y}}+b_2T_{-{\bf y}}) +iq^{1/2}(b_2T_{{\bf
x}}+c_2T_{-{\bf x}}) +iq^{-1/2}(b_3T_{{\bf x}}+c_3T_{-{\bf x}})+\\
+iq^{-1/2}(b_3T_{2{\bf x}+{\bf y}}+c_3T_{-2{\bf x}-{\bf y}})
-(q+q^{-1})(c_1T_{-2{\bf x}}+b_1T_{2{\bf x}}). \end{array}
\end{equation}
The operator $G$ is hermitian if
$d={\bar a},\,\,\,b_1 ={\bar c}_1,\,\,\, -qb_2 ={\bar c}_2,\,\,\, b_3=-q
{\bar c}_3$.
It describes a particle in a magnetic field on the square lattice with a
hopping along ${\bf x}-{\bf y}$ and ${\bf x}+{\bf y}$ diagonals, along $
{\bf x},{\bf y}$ and $2{\bf x}$ bonds and along $2{\bf x}+{\bf y}$ and $3
{\bf x}+{\bf y}$ diagonals.

Although the Hamiltonians (\ref{Gtr}) contains many parameters, the number
of interesting problems is limited since the hopping amplitudes
$t_{\vec\mu}$ must not depend on the magnetic flux (i.e., on $q$). Then
only few cases remain. Among them, there is a triangular lattice:
$b_1=c_1=b_3=c_3=0$, $c_2=b_2=\pm iq^{-1/2}$, $a=d=t\ne 0$ (the sign
depends on parity of $(P-1)/2$: it is "$+$" for $(P-1)/2$ even and "$-$"
otherwise), \begin{equation} \label{H} H=t(T_{{\bf x}+{\bf
y}}+T_{-{\bf x}-{\bf y}})+T_{{\bf x}}+T_{-{\bf x}}+T_{{\bf y}}+T_{-{\bf
y}}\,. \end{equation} Here the triangular lattice is treated as a square
lattice with ${\bf x}+{\bf y}$ diagonals. The square lattice may be
obtained at $t=0$. This case has been considered in \cite{WZ}.

In terms of generators of $U_q(sl_2)$, the Hamiltonian (\ref{H}) is
\begin{equation}
\label{triang4}H=t(A^2+D^2)\pm iq^{-1/2}(q-q^{-1})(CA+BD) \end{equation}
and under the representation
(\ref{reprs})
we obtain the spectral problem
\begin{equation}
\label{triang5}(z^{-1}-tq)\Psi(q^2z)+(zq^{-2}-tq^{-1})\Psi
(q^{-2}z)-(z+z^{-1})\Psi(z)=E\Psi (z)
\end{equation}
How does this equations appears from the original problem? Let us choose a
specific gauge to implement the flux $\Phi /2$ per elementary triangle:
\begin{equation}
\label{triang0}A_{{\bf x}}=-\Phi n_x,\,\,\,\,A_{{\bf y}}=\Phi
n_x,\,\,\,\,A_{ {\bf x}+{\bf y}}=\Phi /2.
\end{equation}
Then, the Schr\"odinger
equation aquires the form
\begin{equation}
\label{triang1}
\begin{array}{c}
e^{-i\Phi n_x}\psi (n_x+1,n_y)+e^{i\Phi (n_x-1) }\psi (n_x-1,n_y)\\
+e^{i\Phi n_x}\psi (n_x,n_y+1)+e^{-i\Phi n_x}\psi (n_x,n_y-1) \\
+t(e^{i\Phi /2}\psi (n_x+1,n_y+1)+e^{-i\Phi /2}\psi
(n_x-1,n_y-1))=E\psi(n_x,n_y)\,.
\end{array}
\end{equation}
Solutions are the Bloch waves:
$$
\psi (n_x,n_y)=e^{ik_xn_x+ik_yn_y}\psi _{n_x}\,, $$
where $\vec k=(k_x$, $k_y$) is the Bloch momentum and $\psi _n$ is $
Q$-periodic discrete function. In these terms (\ref{triang1}) turns into
the following version of the Harper equation: \begin{equation}
\label{triang2}
\begin{array}{c} (e^{ik_x-i\Phi n}+te^{i(k_x+k_y)+i\Phi /2})\psi
_{n+1}+(e^{-ik_x+i\Phi n-i\Phi }+te^{-i(k_x+k_y)-i\Phi /2})\psi _{n-1}+ \\
\\
+2\cos (n\Phi +k_y)\psi _n=E\psi _n.
\end{array}
\end{equation}
The spectrum of the problem has $Q$ bands $E=E_i(k_x,k_y)$, $i=1,...,Q$.
For $k_x=0$, $
k_y=\pi $ (''midband points'' of the spectrum) Eq.(\ref{triang2}) becomes
\begin{equation}
\label{triang3}(q^{-2n}-tq)\psi _{n+1}+(q^{2n-2}-tq^{-1})\psi
_{n-1}-(q^{2n}+q^{-2n})\psi _n=E\psi _n
\end{equation}
with the periodic boundary condition (\ref{10}). This discrete equation is
equivalent to the difference equation (\ref{triang5}).
Applying the results of Sect.3 to (\ref{triang4}),(\ref{triang5}) we obtain
the Bethe equations \begin{equation}
\label{triang51}\frac{z_l(z_l-tq)}{1+tz_l}=\prod_{m=1, \neq l}
^{Q-1}\frac{q^2z_l-z_m}{z_l-q^2z_m}
\end{equation}
and the energies of the mid points of each band: \begin{equation}
E=(q^2-1)\sum_{m=1}^{Q-1}z_m-t(q+q^{-1}). \end{equation}

Harper's equation (\ref{triang2}) for an arbitrary Bloch momentum $\vec k$
may be
obtained from (\ref{triang4}) by using cyclic representations \cite{Skl2}.
These representations
generally has no highest or lowest wieght. The wave functions are not
polynomials in this case. Nevertheless the Bethe Ansatz a kind
of algebraization in
terms of meromorphic functions on higher genus algebraic curves is still
possible. The method
which generalizes the algebraic Bethe Ansatz has been developed in
Ref.\cite {BS} and
applied to the chiral Potts model. Recently, Faddeev and Kashaev \cite{FK}
used this method for the Harper equation on the square lattice to obtain the
Bethe Ansatz equations
for arbitrary $\vec k$.

Simplicity of the spectrum (\ref{Esolv}) of triangular operators considered
in the Sect.3 has a clear interpretation in terms of magnetic translations.
In this case the Hamiltonian allows hopping only
to the South-West, North-East and North-West directions on the lattice.
Therefore, no closed loop trajectory is possible. As a result, the particle
does not feel the magnetic field and the spectrum remains free. \bigskip

\section{Continuum limit. Differential equations} \bigskip

Difference equations of Sect.3 may have different continuum limits. Let us
consider the simplest one, when we assume that $(\Psi(q^{2}z)-
\Psi(q^{-2}z))/(q-q^{-1})$ has a regular limit as $q \rightarrow1$. Let us
set $ q=e^{\hbar}$
and also assume that $q^j=1+{\cal O}(\hbar)$, so $j$ is a fixed integer or
half-integer number (let us note that this condition is not valid for the
periodic difference equations considered in the Sect.4, where
$q^{2j+1}=\pm1$). Then $$ A=1+\hbar S_{3}+\frac{1}{2}\hbar ^{2}S_{3}^{2}
+{\cal O}(\hbar ^{3}) ,\,\,\, D=1-\hbar S_{3}+\frac{1}{2}\hbar
^{2}S_{3}^{2} + {\cal O}(\hbar ^{3}) , $$ $$
B=S_{+}+{\cal O}(\hbar ^{2}) ,\,\,\,\,C=S_{-}+{\cal O}(\hbar ^{2}) $$ as
$\hbar \rightarrow 0$. Here $S_{\pm }$, $S_{3}$ are generators of $
U(sl_{2})$ realized as differential operators (\ref{Dmod}).

The coefficients of $G$ (\ref{G}) may also depend on $\hbar$. Suppose they
are regular at $\hbar =0$. Up to the first order in $\hbar$ one has: $$
a=a^{(0)}+\hbar a^{(1)},\,\,
d=d^{(0)}+\hbar d^{(1)},\,\,
b_{i}=b_{i}^{(0)}+\hbar b_{i}^{(1)},\,\, c_{i}=c_{i}^{(0)}+\hbar
c_{i}^{(1)},\,\,i=1,2,3. $$
In order to get a non-trivial limit one should put $a^{(0)}=d^{(0)}$,
$c_2^{(0)}=-c_3^{(0)}$, $b_2^{(0)}=-b_3^{(0)}$. Then we obtain a general
bilinear form in the $sl_2$ generators (the quantum Euler top
(\ref{Glim})).
Components of the matrix $\alpha_{ij}$ and the vector $\beta_i$ are
$\alpha_{33}=a^{(0)},\,
\alpha_{++}=2b_1^{(0)},\,\alpha_{--}=2c_1^{(0)},\,\alpha_{-,3}=2c_2^{(0)},\,
\alpha_{+,3}=-2b_2^{(0)}
,\,\alpha_{+-}=\alpha_{3,+}=\alpha_{3,-}=0,\,
\beta_{3}=a^{(1)}-d^{(1)},\,\beta_{-}=c_2^{(1)}+c_3^{(1)},\,\beta_{+}
=b_2^{(1)}+b_3^{(1)}$.
Clearly, a global $SL(2)$-rotation
leaves the spectrum invariant. Therefore, in the case of generic position,
one may diagonalize the matrix $\alpha_{ij}$ and reduce the number of
parameters to 4. Indeed, there are 3 momenta of inertia (eigenvalues of
$\alpha _{ij}$) and direction of the "magnetic field" ${\vec \beta}$
(parametrized e.g. by 2 Euler angles). However, the Casimir element ${\vec
S}^{2}$ contributes only to a $c$-number term, so we are left with 4
parameters. In terms of differential operators the adjoint action of the
group amounts to linear fractional transformation. We do not know
$q$-analog of this transformation in the discrete case. An explicit form of
the differential equation which corresponds to the quantum Euler top in the
representation (\ref{Dmod}) is \begin{equation} \label{Lame}
Q_4(z)\frac{d^2}{dz^2}+(Q_2(z)-(j-\frac{1}{2})Q^{\prime}_4(z)) \frac
d{dz}+(\frac{1}{3}j(j-\frac{1}{2})Q_{4}''(z)-jQ^{\prime}_2(z))\,,
\end{equation}
where $Q_k(z)$ are general polinomials in $z$ of degree at most $k$. These
are the only differential operators of second order that have polynomial
eigenfunctions \cite{T3} (see also \cite{T2}).

The quasiclassical version of Bethe equations (\ref{BE1}) describes roots
of the polynomial
eigenfunctions and algebraized part of the spectrum:
\begin{equation}
\label{quasi1}
\frac{Q_{2}(z_{l})-(j-1/2)Q_{4}'(z)}{Q_{4}(z_{l})}=-2 \sum^{2j}_{m=1,m\ne
l} \frac{1}{z_{l}-z_{m} }\,, \end{equation} \begin{equation} \label{quasi2}
E=4b_{1}\sum^{2j}_{n<m}z_nz_m+(b_{2}^{(1)}+b_{3}^{(1)})
\sum^{2j}_{m=1}z_m+const.
\end{equation}
Diagonalization of (\ref{Lame})
in terms of Bethe equations has been proposed in ref.\cite{Ush}.

If $Q_4(z)$ has 4 simple roots, the spectral problem for (\ref{Lame})
is equivalent to the Heun equation
\cite{Erdel}
(the algebraic forms of Lame and Mathieu equations are degenerate
cases of the Heun equation). We, therefore, may call the difference
equation (\ref{Lame}) as $q$-Heun (Mathieu, Lame) algebraic equations.

There is also degenerate (triangular) case. The case (ii) corresponds to
the hypergeometric differential operator
\begin{equation}
\label{Teller}
Q_2(z)\frac{d^2}{dz^2}+(Q_1(z)-(j-1/2)Q_{2}'(z))\frac d{dz}\,. \end{equation}
Its eigenfunctions are classical orthogonal polynomials. The classical
limit of operators of the case (iii) of Sect.3 are first order differential
operators.

At last, let us discuss the continuum limit of the linear form (\ref{L}).
Adopting the similar notation for $\hbar $-dependent coefficients ($
a=a^{(0)}+\hbar a^{(1)}$, etc) as $\hbar \rightarrow 0$ and putting $
a^{(0)}=d^{(0)}$, $b^{(0)}=c^{(0)}=0$ we obtain the operator:
\begin{equation}
\label{Llim}L=2\hbar
^2(2a^{(0)}S_3^2+\sum \beta_iS_i)\,,
\end{equation}
where $\beta_3=a^{(1)}-d^{(1)},\,\beta_{+}=b^{(1)},\,\beta_{-}=c^{(1)}$. As
a differential operator it is:
$$
z^2\frac{d^2}{dz^2}+(-\mu z^2+(1-2j+\nu )z+\lambda )\frac d{dz}+2j\mu z, $$
where $\mu =b^{(1)}/(2a^{(0)}), \nu =(a^{(1)}-d^{(1)})/(2a^{(0)}),
\lambda=c^{(1)}/(2a^{(0)})$. After the change of variable $z=e^{-x}$ and a
''gauge'' transformation eliminating the term with first derivative, this
operator aquires the form $\frac{d^2}{dx^2}-V(x)$ with $$
V(x)=\frac{1}{4}(\mu ^2e^{-2x}+\lambda ^2e^{2x})-\frac 12(\mu (2j+1+\nu
)e^{-x}+\lambda (2j+1-\nu )e^x)\,.
$$
If $c^{(1)}=0$ or $b^{(1)}=0$ it is the Morse potential. \bigskip

\section{Difference equations solvable in symmetric Laurent
and Askey-Wilson polynomials}\bigskip

In this section we describe a related family of difference operators whith
eigenfunctions given by {\it symmetric Laurent polynomials} in one variable
$y$, i.e., invariant under the inversion $y\rightarrow y^{-1}$.
Differential operators of this kind may be obtained from (\ref{Lame}) and
(\ref{Teller}) by the change of variable $z=(y+y^{-1})/2$ (that is the
change $z=\cos \theta $, $e^{i\theta }=y$ which turns Legendre polynomials
into spherical harmonics). In the case of difference equations, there is no
direct $q$-analog of "changes of variables". We must use another operator
algebra.

The proper operator algebra has been obtained in the paper \cite{GZ} as a
certain degenerate case of the Sklyanin algebra \cite{Skl2}:
\begin{equation}\label{B2}\begin{array}{c} \tilde D\tilde C=q\tilde C\tilde
D,\,\,\,\,\,\,\,\,\,\, \tilde C\tilde A=q\tilde A\tilde C,\\ \\ \tilde
A\tilde B-q\tilde B\tilde A=q\tilde D\tilde B-\tilde B\tilde D=
-(1/4)(q^{2}-q^{-2})(\tilde D\tilde C-\tilde C\tilde A),\\ \\
\phantom{a}[\tilde A,\tilde D]=(1/4)(q-q^{-1})^{3}\tilde C^{2},\\ \\
\phantom{a}[\tilde B,\tilde C]=
\displaystyle{\frac{ \tilde A^{2}-\tilde D^{2} }{q-q^{-1}}}\,.
\end{array}\end{equation}
It can be realized by the shift operators (\ref{inv1}): \begin{equation}
\label{B1}\begin{array}{l}
\tilde A=\displaystyle{\frac{q^{-j}}{y-y^{-1}}} (yT_{+}-y^{-1}T_{-}),\,\,\,\,
\tilde D=\displaystyle{\frac{q^{j}}{y-y^{-1}}}(-y^{-1}T_{+}+yT_{-}),\\ \\
\tilde B=\displaystyle{\frac{1}{2(q-q^{-1})(y-y^{-1})}}
(-(q^{-2j}y^{2}+q)(q^{2j-1}y^{-2}+1)T_{+}+ \\ \\
+(q^{-2j}y^{-2}+q)(q^{2j-1}y^{2}+1)T_{-}),\\ \\ \tilde
C=\displaystyle{\frac{2}{(q-q^{-1})(y-y^{-1})}}(T_{+}-T_{-}).
\end{array}\end{equation}
The representation space of this algebra is spanned by $y^{k}+y^{-k}$,
$k=$0,\,1,...,\,2$j$ (symmetric Laurent polynomials). The standard quantum
algebra $U_q(sl_2)$ can be obtained as a contraction of (\ref{B2}): $\tilde
C=\epsilon ^{2}C$, $\tilde A=\epsilon A$, $\tilde D=\epsilon D$, $\tilde
B=B$, $\epsilon \rightarrow 0$. There are two central elements:
\begin{equation}\label{B3}
\tilde \Omega _{0} =\tilde A\tilde
D+\frac{1}{4q}(q-q^{-1})^{2} \tilde C^{2}, \end{equation}
\begin{equation}\label{B4}
\tilde \Omega _{1}=\frac{ q^{-1}\tilde A^{2}+q\tilde D^{2} }
{(q-q^{-1})^{2} }+\tilde B\tilde C+\frac{1}{2}(q+q^{-1})\tilde C^{2}.
\end{equation}
In the continuum limit one obtains $U(sl_2)$ in the representaion
(\ref{reprs}), where
$z=(y+y^{-1})/2$.
Homogeneous bilinear (and linear)
forms in $\tilde A,\,\tilde B,\,\tilde C,\,\tilde D$ give rise to a class
of difference equations partially solvable in symmetric Laurent
polynomials.

The linear form (\ref{L}) (with $\tilde A,\,\tilde B,\,\tilde C,\, \tilde
D$ in place of $A,\,B,\,C,\,D$) gives the operator \begin{equation}
\label{B001}
\frac{ \tilde a(q^{-j}y)}{2(y-y^{-1})}T_{+}- \frac{ \tilde
a(q^{-j}y^{-1})}{2(y-y^{-1})}T_{-}, \end{equation}
where
\begin{equation} \label{B002}
\tilde a(y)=-b(y^{2}+y^{-2})+2ay-2dy^{-1}+4c-b(q+q^{-1}). \end{equation}
If $b\ne 0$ and $c\ne 0$ the algebraic eigenfunctions of (\ref{B001}) are
given by
\begin{equation} \label{B10}
\Psi (y)=y^{-2j}\prod _{l=1}^{2j}(y-y_{l})(y-y_{l}^{-1}) \end{equation}
and $y_{l}$'s satisfy the Bethe equations \begin{equation} \label{B003}
\frac{ \tilde a(q^{-j}y_{k}^{-1})(y_{k}-qy_{k}^{-1})} {\tilde
a(q^{-j}y_{k})(qy_{k}-y_{k}^{-1})}= -\prod _{l=1, l\ne k}^{2j}\frac{
(qy_{k}-y_{l})(qy_{k}-y_{l}^{-1})} {(y_{k}-qy_{l})(y_{k}-qy_{l}^{-1})},
\end{equation}
with the eigenvalues being expressed through $y_{l}$'s as follows:
\begin{equation} \label{B004}
E=aq^j +dq^{-j}-\frac{1}{2}b(q-q^{-1})\sum _{l=1}^{2j}(y_{l}+y_{l}^{-1})
\end{equation}
(compare with (\ref{linenergy1}).

The general bilinear form (\ref{G}) in $\tilde A,\,\tilde B,\,\tilde C,
\,\tilde D$ gives the difference operator \begin{equation} \label{B6}
G=A(y)(T_{+}^{2}-1)+A(y^{-1})(T_{-}^{2}-1)+W(y), \end{equation}
where
\begin{equation} \label{B7}
A(y)=\frac{ \sum _{s=-4}^{4}a_{s}q^{-js}y^{s} } {(y-y^{-1})(qy-q^{-1}y^{-1}) },
\end{equation}
\begin{equation} \label{B8} \begin{array}{c} W(y)=\frac{1}{4}(q^{2j}-q^{-2j})
\left( b_{1}(q^{2j-1}-q^{-2j+1})(y^{2}+y^{-2})+ \right.\\ \\
\left.+2(b_{2}q^{j}+b_{3}q^{-j})(y+y^{-1})\right), \end{array}
\end{equation}
\medskip
and
\begin{equation} \label{B9} \begin{array}{c} a_0 =4c_1
+\frac{1}{2}b_{1}(q^{2}+q^{-2}+1),\,\, a_1 =2q(c_2
+\frac{1}{4}(q^{-2}b_{2}-(q+q^{-1})b_{3})),\\ \\
a_{-1}=-2q^{-1}(c_{3}+\frac{1}{4}(q^{2}b_{3}-(q+q^{-1})b_{2})),\,\,
a_{2}=qa+\frac{1}{4}q(q+q^{-1})^{2}b_{1},\\ \\
a_{-2}=q^{-1}d+\frac{1}{4}q^{-1}(q+q^{-1})^{2}b_{1},\\ \\
a_{3}=-\frac{1}{2}qb_{3},\,\,
a_{-3}=\frac{1}{2}q^{-1}b_{2},\,\,
a_{4}=\frac{1}{4}q^{2}b_{1},\,\,
a_{-4}=\frac{1}{4}q^{-2}b_{1}.
\end{array} \end{equation}

In the generic case (at least one of $b_{i}$'s is not 0) the algebraic
eigenfunctions of (\ref{B6}) are given by (\ref{B10}), where $y_{l}$'s obey
the following system of Bethe equations: \begin{equation} \label{B11}
\frac{ \sum _{s=-4}^{4} a_{s}q^{-js}y_{k}^{-s} } { \sum _{s=-4}^{4}
a_{s}q^{-js}y_{k}^{s} }= \prod _{l=1, l\ne k}^{2j}\frac{
(q^{2}y_{k}-y_{l})(q^{2}y_{k}y_{l}-1)} {
(y_{k}-q^{2}y_{l})(y_{k}y_{l}-q^{2}) }. \end{equation}
The eigenvalues are
\begin{equation} \label{B12}
\begin{array}{c}
E=\frac{1}{4}b_{1}(q-q^{-1})(q^{2}-q^{-2}) \displaystyle{\sum _{l<m}^{2j}}
(y_{l}+y_{l}^{-1})(y_{m}+y_{m}^{-1})- \\
\\-\frac{1}{2}(q-q^{-1})(b_{2}q^{-j+1}+b_{3}q^{j-1}) \displaystyle{\sum
_{l=1}^{2j}}
(y_{l}+y_{l}^{-1})+(q^{2j}-q^{-2j})(a-d)-\\ \\
-\frac{1}{4}b_{1}(q+q^{-1})(q^{2j}-q^{-2j})^{2}+
\frac{1}{2}jb_{1}(q-q^{-1})(q^{2}-q^{-2}). \end{array} \end{equation}

The "exactly solvable" case $b_1 =b_2 =b_3 =0$ when the operator is
triangular is of particular
importance. Its eigenfunctions are known as {\it Askey-Wilson polynomials}
\cite{AW0,AW}. In this case $W(y)=0$ in (\ref{B6}). For references we
recall the conventional form of the difference equation for Askey-Wilson
polynomials:
\begin{equation}
\label{B13}\begin{array}{c}
A(y)(\Psi _{n}(q^{2}y)-\Psi _{n}(y))+A(y^{-1}) (\Psi _{n}(q^{-2}y)-\Psi
_{n}(y))=\\ \\
=(q^{-2n}-1)(1-q^{2n-2}w_1 w_2 w_3 w_4)\Psi _{n}(y) \end{array} \end{equation}
with
\begin{equation} \label{B14}
A(y)=\frac{ \prod _{\alpha =1}^{4}(1-w_{\alpha }y) } {
(1-y^{2})(1-q^{2}y^{2}) },
\end{equation}
where $w_{i}$ are independent parameters\footnote{They are connected with
the previous ones in an obvious way; in particular, $w_1 w_2 w_3 w_4
=q^{2-4j}a/d$.}. The zeros $y_{l}$, $y_{l}^{-1}$ of the Askey-Wilson
polynomials satisfy the system of Bethe equations (\ref{B11}).
In the notation of eqs.(\ref{B13}),(\ref{B14}) they are \begin{equation}
\label{B16} \prod _{\alpha =1}^{4}\frac{y_{k}-w_{\alpha
}}{w_{\alpha }y_{k}-1}= \prod _{l=1, l\ne k}^{n}\frac{ (q^{2}y_{k}-y_{l})
(q^{2}y_{k}y_{l}-1)}{(y_{k}-q^{2}y_{l})(y_{k}y_{l}-q^{2}) }\,. \end{equation}
The form of these equations suggests to ask for an interpretation of the
Askey-Wilson polynomials in terms of integrable spin chains with
boundaries.

Setting $y=e^{2x}$, $w_1 =q^{\alpha }$, $w_2 =q^{\beta }$, $w_3 =-q^{\gamma
}$, $w_4 =-q^{\delta }$, one readily finds that in the continuum limit the
difference operator in the l.h.s. of (\ref{B13}) turns into
\begin{equation} \label{B17}
\frac{d^{2}}{dx^{2}}-\frac{1}{\sinh (2x)}\left ( 2(\gamma +\delta -\alpha
-\beta )+
(2-\alpha -\beta -\gamma -\delta )\cosh (2x) \right ) \frac{d}{dx}\,.
\end{equation}
Note that in the continuum
limit the number of independent parameters is reduced to 2. Eliminating the
first derivative term by means of a suitable "gauge" transformation, one
arrives at the Schr\"odinger equation for the
generalized P\"oschl-Teller potential. The same potential appears from
(\ref{Teller}) after a suitable change of variable and a "gauge"
transformation. The solutions of the P\"oschl-Teller equation are Jacobi
polynomials.

A particular case of Askey-Wilson polynoms $w_1 =-w_3$, $w_2 =-w_4=q$ is
Rogers-Askey-Ismail (or $q$-Gegenbauer) polynomials \cite{AW}. The
corresponding spectral problem (\ref{B13}) in this case have a simple
multivariable generalization. It is the $q$-analog (or relativistic
extension) of the Calogero-Moser system \cite{Ruij}.

The interrelation between big $q$-Jacobi and Askey-Wilson polynoms is
two-fold. First, the former can be obtained \cite{K} under a certain
scaling limit from the latter, although no parameter is lost in the course
of this limit. The two families
appear to be equivalent. Furthemore, there exists a similarity
transformation (isospectral transformation) connecting the two families
(see the end of the next section). One can interpret this transformation as
change of variables in the difference equation. \bigskip

\section{Conclusion and discusion}
\bigskip

In this paper we have attempted to classify the difference and discrete
operators that preserve
a space of polynomials. These operators may be represented as bilinear (or
linear)
forms of the
generators of $U_{q}(sl_{2})$ (q-analogs of the quantum Euler top).
Difference and discrete
equations appearing as spectral problems for these operators have
polynomial solutions. Roots
of polynomials and eigenvalues are given by the Bethe Ansatz equations.
Generally polynomial
eigenfunctions cover only a part of the spectrum. Sometimes all the
spectrum is algebraized. Differential equations having polynomial solutions
appear in a certain continuous limit of the difference equations. They have
been classified previously in Refs. \cite{T2,Ush,Sh}.

Some periodic difference equations with an incommensurate period have a
peculiar
multifractal spectrum (Harper's equation (\ref{5}) is an example). The
Bethe Ansatz equations may be helpfull for a statistical description of the
multifractality.

Algebraized difference equations have an intimate relation with integrable
quantum and classical nonlinear systems. Below we discuss a particular
aspect of this relation.

The Heun (Mathieu, Lame) operator
(\ref{Lame}) written in the {\it algebraic form} may be transformed to the
Schr\"odinger-type) operator
\begin{equation}
\label{trans}H=\frac{d^2}{dx^2}+V(x)
\end{equation}
by the change of variable
\begin{equation}
\label{change}x=\int\limits^z\frac{dy}{\sqrt{Q_4(y)}} \end{equation}
and a subsequent "gauge" transformation. This is the {\it
transcendental form} of the Heun (Mathieu, Lame) operator. The potential
$V(x)$ is a certain
elliptic function \footnote{For its explicit form see \cite{ZU},
\cite{Ush}.}. In the
case of Lame equation it
is the Weierstrass function: $V(x)=n(n+1)\wp (x)$.

Correspondence between solutions of the
algebraic and transcendental forms of the equations is peculiar. The
potential $V(x)$ is a periodic
function, so the spectrum of the operator (\ref{trans}) has band structure.
Some polynomial eigenfunctions of the algebraic form (\ref{Lame}) describe
edges of the bands, others correspond to unnormalizable eigenfunctions
\cite{T2},\cite{Olver}. Both operators
(\ref{trans}) and (\ref{Lame}) may have also other eigenfunctions which do
not correspond to each other.

An important property of the potential of the Lame equation in the
transcendental
form is
its direct generalization to the many body case:
$$-\sum_{i}\partial^2_i+\sum_{i>j}n(n+1)\wp (x_i-x_j).$$ This is the
elliptic version of the celebrated Calogero-Moser model
\cite{OP}. Than several questions arise:

(i) Is there a many-body version of the Lame equation in the algebraic
form? If yes, it must
be a new quantum integrable theory, equivalent to the Calogero-Moser
model, and having an explicit quantum group symmetry;

(ii) Do the general
Heun (Mathieu) operator in algebraic (\ref{Lame}) or transcendental
\ref{trans} form have a many-body generalization?

The same questions aplly to the $q$-deformed equations:

(iii) The operator (\ref{diffeq}) is the $q$-deformation of the {\it
algebraic form} (\ref{Lame}) of the Heun operator. What it is a $q$-analog
of its transcendental form (\ref{trans}) and the transformation
(\ref{change})?

(iv) The $q$-deformation of the
Calogero-Moser hamiltonian is known. It is the Macdonald operator
\cite{Macdonald}. What is the $q$-deformed many-body version of the general
Heun operator? What is the algebraic form of this many-body theory?

Here we may give a partial answer to the question (iii).
We suggest
that the difference equation for the Askey-Wilson polynomials \cite{AW}
(see eq.(\ref{B13}) ) is the $q$-deformation of the trigonometric limit
of the Lame operator in the
transcendental form. Under some {\it a similarity transformation} it
becomes equivalent to the triangular operator of the case (ii) of
Sect.3. The explicit form of this transformation is quite complicated. Its
existence follows from the
representation theory of the quadratic algebra (\ref{algH}) developed in
the paper \cite{GZh1}.
We plan to address these questions elsewhere. \bigskip

\section*{Acknowledgements}\medskip
We are grateful to A.Gorsky, J.Shnittger, A.Turbiner for useful discussions
and C.Zachos for bringing Refs. \cite{Fairlie},\cite{Zachos},\cite{Vinet}
to our
attention. The work of A.Z. was partially supported by
 grant 93-02-14365 of the Russian Foundation of Fundamental Research,
 by ISF grant MGK000, and by ISTC grant 015. P.W. thanks Universite
 de Paris	and
Yukawa Institute for Theoretical Physics for the
 hospitality where this work has been done. This workWe are grateful to
A.Gorsky, J.Shnittger, A.Turbiner for useful
discussions and C.Zachos for bringing Refs.
\cite{Fairlie},\cite{Zachos},\cite{Vinet}
to our
attention. The work of A.Z.  was partially supported by grant 93-02-14365
of the Russian Foundation of Fundamental Research, by ISF grant MGK000,
and by ISTC grant 015. P.W. thanks Laboratoire de Physique Theorique et
Hautes Energies
 at  Universite de Paris VI and
Yukawa Institute for Theoretical Physics in Kyoto for the hospitality where
this
work has been done. P.W. was supported primerely by the MRSEC Program of the
 NSF under Award Number DMR-9400379.

\bigskip

\appendix \section*{Appendix A}
\def\theequation{A\arabic{equation}}
\setcounter{equation}{0}

Let us show that linear difference operators of the form
(\ref{1}) have $(n+1)$-dimensional invariant subspace of polynomials
$Pol_n$ iff the coefficients $a(z),\,d(z),\,v(z)$ are certain Laurent
polynomials of order 2 with 7 independent parameters. We assume that $n>1$.
Let us denote
$f_1(z)=a(z),\,f_2(z)=v(z),\,f_3(z)=d(z)$, and consider the decomposition
of the coefficients
$$
f_{i}(z)=f_{i}^{(0)}+f_{i}^{(+)}(z)+f_{i}^{(-)}(z), $$
into a constant $f_{i}^{(0)}$, strictly positive $f_{i}^{(+)}$ and negative
$f_{i}^{(-)}$ degrees of
$z$.
Similarly, the operator
$\Delta =\Delta ^{(0)}+\Delta ^{(+)}+\Delta ^{(-)}$, where $$\Delta \equiv
f_{1}T_{+}+f_{2}+f_{3}T_{-}$$ is decomposed into a raising (lowering) parts
$\Delta ^{(+)}$($\Delta ^{(-)}$ and a diagonal operator $\Delta ^{(0)}$.
Clearly, $\Delta $ leaves $Pol_n$ invariant if and only if both $\Delta
^{(+)}$ and $\Delta ^{(-)}$ do so separately. Consider, say, $\Delta
^{(+)}$. To preserve the space
$Pol_n$ it must annihilate $z^n$ and
transform each vector $z^k$ ($k<n$) to a linear combination of
$z^{k+1}$, $z^{k+2}$,...,$z^{n}$. This gives a number of conditions, the
first three of them are \begin{equation} \label{A1} \begin{array}{c}
q^{n}f_{1}^{(+)}(z)+f_{2}^{(+)}(z)+q^{-n}f_{3}^{(+)}(z)=0, \\ \\
q^{n-1}f_{1}^{(+)}(z)+f_{2}^{(+)}(z)+q^{-n+1}f_{3}^{(+)}(z)= A_{1}^{(+)}z,
\\ \\
q^{n-2}f_{1}^{(+)}(z)+f_{2}^{(+)}(z)+q^{-n+2}f_{3}^{(+)}(z)=
A_{2}^{(+)}z^{2}+A_{3}^{(+)}z\,,
\end{array}
\end{equation}
where $A_{i}^{(+)}$ are arbitrary coefficients. This is a system of linear
equations for $f_{i}^{(+)}$ with non-zero determinant (it is equal to
$(q-q^{-1})(q+q^{-1}-2)$). Therefore, $f_{i}^{(+)}(z)$ are uniquely
determined from (\ref{A1}) to be linear combinations of $z$ and $z^2$ with
3 independent parameters $A_{i}^{(+)}$. All the other conditions are then
automatically satisfied. Similar arguments applyed to $\Delta ^{(-)}$ allow
one to find $f_{i}^{(-)}(z)$ as linear combinations of $z^{-1}$ and
$z^{-2}$ parametrized by 3 independent constants $A_{i}^{(-)}$.

The total number of independent parameters is 9 ($A_{i}^{(\pm )}$,
$f_{i}^{(0)}$; $i=$1,\,2,\,3), i.e., the linear space of 2-nd order
difference operators preserving the space of polynomials is 9-dimensional.
Two of these parameters correspond to the constant term and the common
factor in (\ref{inv1}), so there are 7 essential parameters (not including
$q$).

Calculating $f_{i}^{(\pm )}(z)$ explicitly, one finds that $\Delta $
(\ref{inv1}) coincides (after obvious changes $n\rightarrow 2j$,
$q\rightarrow q^2$) with the operator (\ref{diffeq}) obtained by means of
$U_q(sl_2)$. \bigskip

\appendix\section*{ Appendix B}
\def\theequation{B\arabic{equation}}
\setcounter{equation}{0}
\bigskip

{\bf B1.}\,\,There are other weight representations of $U_q(sl_2)$ by the
"shift" operators $T_{\pm }$ (see (\ref{inv1})). One of them is
\begin{equation}
\label{rep1}
\begin{array}{l}
A=q^{-j}T_{+},\,\,\,\,\,\,D=q^{j}T_{-},\\ \\
B=(q-q^{-1})^{-1}z(-q^{-j}T_{+}+q^{j}-q^{-j-1}+q^{j-1}T_{-}), \\ \\
C=(q-q^{-1})^{-1}z^{-1}(q^{-j}T_{+}+q(q^{j}-q^{-j-1})-q^{j+1}T_{-})\,.
\end{array}
\end{equation}
The representation space again consists of polynomials of degree $2j$ and
the value of the Casimir operator is the same as in (\ref{casimir2}). The
classical limit of (\ref{rep1}) is the same as that of (\ref{reprs}).
Bearing in mind applications to 2-nd order difference equations,
(\ref{reprs}) is more convenient than (\ref{rep1}) because any homogeneous
bilinear form in $A$, $B$, $C$, $D$ realized as in (\ref{reprs}) becomes a
2-nd order difference operator, whereas a general bilinear form in the
representation (\ref{rep1}) gives an operator of fourth order.

Another representation is realized in the space of Taylor series in $z$
\cite{Vinet,Vays}:
\begin{equation}
\label{rep2} \begin{array}{l} A=q^{j}T_{-},\,\,\,\,\,\,D=q^{-j}T_{+}, \\ \\
B=(q-q^{-1})^{-2}z^{-1}(q^{-2j-1}(1-T_{+}^{2})+ q^{2j+1}(1-T_{-}^{2})),\\
\\
C=z\,.
\end{array}
\end{equation}
There is an invariant subspace spanned by $z^k$ with $k>2j$. The induced
representation in the (finite-dimensional) factorspace is equivalent to the
spin $j$ representation. The representation (\ref{rep2}) (in a slightly
different form) was used in \cite{Vinet} for discretizing the Schr\"odinger
equation with $x^{-2}$-potential. \medskip

{\bf B2.}\,\,
Algebraized difference equations may be obtained from any weight
representations of $q$-deformations of $sl_2$ other than $U_q(sl_2)$
(\ref{ABCD}).

One of them is generated by
$J_{\pm }$, $J_{0}$ \cite{Witten}:
\begin{equation}
\label{alg1}
\begin{array}{l}
q^{-1}J_{+}J_{0}-qJ_{0}J_{+}=-J_{+}, \,\,
qJ_{-}J_{0}-q^{-1}J_{0}J_{-}=J_{-}, \\ \\
q^{-2}J_{+}J_{-}-q^{2}J_{-}J_{+}=(q+q^{-1})J_{0}. \end{array}
\end{equation}
This algebra covers a part of $U_q(sl_2)$: \begin{equation}
\label{alg2}
\begin{array}{l}
J_{+}=BD=(q-q^{-1})^{-1}q^{j}z(q^{2j}T_{-}^{2}-q^{-2j}), \\ \\
J_{-}=CD=(q-q^{-1})^{-1}q^{j}z^{-1}(1-T_{-}^{2}), \\ \\
J_{0}=(q-q^{-1})^{-1}(1-wD^2)=(q-q^{-1})^{-1}\left(1-
\displaystyle{\frac{q^{4j+1}+q^{-1}}
{q+q^{-1}}}T_{-}^{2}\right)\,,
\end{array}
\end{equation}
where $$ w=(q+q^{-1})^{-1}(q-q^{-1})^{2}\Omega \,. $$ Another "half" of the
algebra may be obtained by the authomorphism: $A\rightarrow D$,
$D\rightarrow A$, $B\rightarrow C$, $C\rightarrow B$.

This algebra has been used in the papers \cite{OT,T2,T3} to construct
difference equations having polinomial solutions. A class of difference
equations obtained from this algebra is of the form $G \Psi=\varepsilon
T_{\pm}^{2} \Psi$, where $G$ is the invertable Jacobi operator (\ref{G}).
Generally these equations do not correspond to any hermitian operator.

All other deformations of $sl_2$
have a similar embedding into $U_q(sl_2)$. More information about different
$q$-deformed algebras and
interrelations between them may be found in \cite{Zachos}. Some particular
examples were studied in \cite{Odes}, \cite{Woron}, \cite{Fairlie}.
Another algebra of difference operators which preserves the space of
symmetric Laurent polynomials is considered in Sect.6. 

\end{document}